\documentclass[10pt,twocolumn,a4paper]{IEEEtran}

\usepackage{amsmath}
\usepackage{url}
\usepackage{xcolor}
\usepackage[utf8]{inputenc}
\usepackage[algo2e,ruled,vlined]{algorithm2e}
\usepackage{titlesec}
\usepackage{float}
\usepackage{amssymb}
\usepackage{graphicx}
\usepackage{authblk}

\titleformat{\subsection}[runin]
       {\normalfont\bfseries}
       {\thesubsection}
       {0.5em}
       {}
       [.]
\title{Giga-voxel multidimensional fluorescence imaging combining single-pixel detection and data fusion}

\author[1,2,*,$\dagger$]{F. Soldevila}
\author[2,*]{A. Lenz}
\author[3,4]{A. Ghezzi}
\author[4]{A. Farina}
\author[3,5]{C. D'Andrea}
\author[2]{E. Tajahuerce}

\affil[1]{Laboratoire Kastler Brossel, École Normale Supérieure – Paris Sciences et Lettres (PSL) Research University, Sorbonne Université, Centre National de la Recherche Scientifique (CNRS) UMR 8552, Collège de France, 24 rue Lhomond, 75005 Paris, France.}
\affil[2]{GROC-UJI, Institute of New Imaging Technologies (INIT), Universitat Jaume I, 12071, Avda. Sos Baynat, s/n, Castelló, Spain.}
\affil[3]{Politecnico di Milano, Dipartimento di Fisica, Piazza L. da Vinci 32, 20133 Milano, Italy.}
\affil[4]{Consiglio Nazionale delle Ricerche, Istituto di Fotonica e Nanotecnologie, Piazza L. da Vinci 32, 20133 Milano, Italy.}
\affil[5]{Istituto Italiano di Tecnologia, Center for Nano Science and Technology, via Pascoli 70/3, 20133 Milano, Italy.}
\affil[*]{These authors contributed equally to this work}
\affil[$\dagger$]{Corresponding author: soltfern@gmail.com}

\begin{document}
\twocolumn[
\begin{@twocolumnfalse}
\maketitle

\begin{abstract}
Time-resolved fluorescence imaging is a key tool in biomedical applications, as it allows to non-invasively obtain functional and structural information. However, the big amount of collected data introduces challenges in both acquisition speed and processing needs. Here, we introduce a novel technique that allows to reconstruct a Giga-voxel 4D hypercube in a fast manner while only measuring 0.03\% of the information. The system combines two single-pixel cameras and a conventional 2D array detector working in parallel. Data fusion techniques are introduced to combine the individual 2D \& 3D projections acquired by each sensor in the final high-resolution 4D hypercube, which can be used to identify different fluorophore species by their spectral and temporal signatures. 
\end{abstract}
\end{@twocolumnfalse}
]

During the last few decades, the amount of data being collected by optical systems has been growing at an exponential rate. Nowadays, bio-imaging researchers are not only interested in obtaining high-resolution (over millions of pixels) images, but also in measuring additional physical properties of light, such as polarization, wavelength, and fluorescence lifetimes \cite{orth_gigapixel_2015,wang_single-shot_2020}. Furthermore, state of the art biological research spans from the study of thin microscopic 2D samples to full organisms \textit{in vivo}, thus requiring 3D, fast, and highly-dimensional imaging systems \cite{prevedel_simultaneous_2014,fan_video-rate_2019}.

This increase in the amount of acquired data presents several challenges. First, imaging systems need to be designed with the capability to sense not only light intensity, but also other physical parameters (wavelength, polarization, time-resolved decays on the ps timescale, etc.) and to operate in real-time. Current detector and electronics technology are limited mainly by the fact that detectors are sensitive only to the intensity of light and the technical limitations when building a sensor. Manufacturing places a bound in the number of pixels that can be fitted in a given sensor size, and working conditions (cooling, power supply, etc.) generate trade-offs between the number of physical parameters which can be measured and any combination of frame-rate, pixel size, sensitivity, quantum efficiency, and/or pixel number. Another main challenge is that, even when multidimensional systems can be built with adequate specifications, the amount of data generated tends to be so big that bottlenecks in transmission, storage, and computational power limit the capability of such systems to perform in real-time \cite{rueden_visualization_2007}.

Recently, single-pixel (SP) imaging systems have been proposed as a way to tackle some of these limitations. SP cameras operate with a single bucket detector and a spatial light modulator (SLM). The SLM is used to sample the scene by using coded masks, and the total intensity of the superposition among the masks and the scene is measured with a detector using just one pixel \cite{edgar_principles_2018}. In contrast with a conventional camera, which uses millions of pixels to provide sharp images, SP imaging systems shift the spatial sampling process to the SLM. By doing this, simple but extremely specialized detectors can be used, which allow to build very efficient multidimensional systems \cite{radwell_single-pixel_2014,rousset_time-resolved_2018,soldevila_phase_2018}. Moreover, image recovery in SP systems is very well suited to signal processing techniques, such as compressive sensing or machine learning \cite{duarte_single-pixel_2008,jiang_imaging_2020}, which help alleviate the aforementioned data processing hurdles. However, SP systems are not exempt of limitations. As the SLM needs to generate multiple masks to sample the scene, SP systems are sequential in nature, and thus are bounded by a trade-off between spatial resolution and frame-rate.

Using a different approach, data fusion (DF) techniques aim to combine any number of individual datasets into one single dataset that provides richer information than any one of the starting ones. In the same way humans merge information from sight, smell, or touch to determine if it is safe to eat some food, multidimensional data fusion systems are able to provide novel insights on sample characteristics from a combined view of multispectral, time-resolved, and/or polarimetric views of the scene. Historically, the main field of application of DF has been remote sensing, where satellite design imposes hard constraints on the energy consumption, bandwidth, and number and size of detectors \cite{zhang_multi-source_2010,khaleghi_multisensor_2013}. Given these limitations, it is quite normal to have multiple sensors, each one being sensitive to a different spectral range or to the polarization state of light. After capturing all the data, the fusion procedure helps to obtain rich chemical and morphological information about the surface. With the same spirit, there has been a recent spark of interest on DF in the life sciences, as merging information from different imaging modalities has proven to give insights that individual sources cannot provide \cite{kessler_image_2006,smith_two_2012,van_de_plas_image_2015,fatima_enhanced-resolution_2020}.

In this letter we present a novel technique that combines both the SP and DF paradigms. By doing so, it allows to capture high spatial resolution, multispectral, and time-resolved fluorescence images. Both spectral features and fluorescence lifetimes provide fundamental insights about the photophysical processes of many different samples. In particular, emission spectra allow to distinguish among different chemical species, while fluorescence lifetimes, being strongly dependent on the fluorophores microenvironment, provide useful functional information (e.g. pH, temperature, energy transfer, etc.). The capture process is achieved while still using simple detectors that individually gather information about a reduced number of dimensions (space, time, wavelength). Our system relies on the combined use of three different sensors: two SP cameras capturing multispectral and time-resolved information, and a conventional array detector capturing high spatial resolution images. After the measurement process, DF techniques are introduced to combine the individual 2D/3D projections acquired in parallel by each sensor in the final 4D hypercube. This provides an efficient system that is not bound by bandwidth and storage limitations, as each individual sensor only measures a small fraction of information. Furthermore, the DF procedure is done by simply solving a regularized inverse problem via gradient descent without the requirement of the calculation of the Hessian, which typically entails memory limitations.

\begin{figure}[hbt!]
\centering
\includegraphics[width=\linewidth]{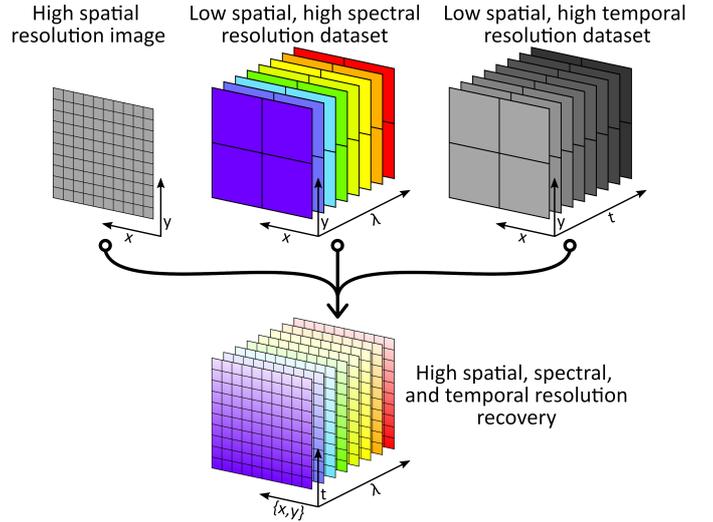}
\caption{ {\bf Spatio-temporal-spectral data fusion framework.} A CMOS camera acquires a high spatial resolution image with neither temporal nor spectral resolution. A SP multispectral camera acquires a low spatial, but high spectral resolution datacube, using a spectrometer as its detector. Last, an additional SP camera measures a low spatial, but high temporal resolution datacube, using a fast bucket detector. All three datasets are combined via regularization to obtain a 4D high resolution spatial, temporal, and spectral hypercube.}
\label{fig:fusion}
\end{figure}

Our system combines the images obtained with two SP cameras with an image obtained with a CMOS camera (see Fig. \ref{fig:fusion}). Individually, each SP camera provides either multispectral or time-resolved images with a low spatial resolution, while the CMOS sensor captures a high spatial resolution image of the sample, but neither spectral nor time-resolved. The DF procedure makes it possible to retain the SP benefits of using simple specialized detectors while still obtaining high spatial resolution images. This allows to acquire full 4D reconstructions (x, y, wavelength, time) of a fluorescent sample with multiple fluorophore species.

We model our system in the following way. For each camera, we can formulate a forward model that represents the acquisition of a projection of the 4D hypercube ($\mathbf{x}$) over several dimensions. For example, for the single CMOS image we have $\mathbf{y}_{cmos} = S \cdot T \cdot \mathbf{x}$, where $S$ and $T$ represent the spectral and temporal integration operators (i.e.  $S$ and $T$, in combination, project the 4D hypercube over the 2D space). In the same way, we can define forward models for both the spectral and time-resolved SP cameras. For the spectral camera we have $\mathbf{y}_{spectral} = T \cdot R_L \cdot \mathbf{x}$, where $R_L$ is a downsampling operator in the spatial domain (as the SP cameras acquire low spatial resolution images). Last, for the time-resolved camera we have $\mathbf{y}_{temporal} = S \cdot R_L \cdot \mathbf{x}$. Given $\mathbf{y}_{cmos}$, $\mathbf{y}_{spectral}$, and $\mathbf{y}_{temporal}$, the problem then resides on finding an estimation of the hypercube, ${\mathbf{\hat{x}}}$, that is compatible with all the individual measurements. To do so, we formulate the following minimization problem:
\begin{equation}
\mathbf{\hat{x}} = \underset{\mathbf{x}}{\text{arg min }} F(\mathbf{x}) 
\label{eq:eq1}
\end{equation}
\begin{equation}
\begin{split}
F(\mathbf{x}) &= \frac{1}{2} \| S T \mathbf{x} - \mathbf{y}_{cmos} \|_{2}^{2} 
+ \frac{1}{2} \alpha \| R_L S \mathbf{x} - \mathbf{y}_{temporal}\|_{2}^{2} + \\
			&+ \frac{1}{2} \beta \| R_L T \mathbf{x} - \mathbf{y}_{spectral} \|_{2}^{2}.
\label{eq:eq2}
\end{split}
\end{equation}

The first term in Eq. \ref{eq:eq2} minimizes the difference between the measurements obtained with the CMOS camera and the projection of the 4D hypercube over the 2D space. The second term minimizes the difference between the time-resolved SP measurements and the projection of the 4D hypercube over a low-resolution 3D space (x, y, time). Last, the third term minimizes the difference between the SP multispectral measurements and the projection of the 4D hypercube over a low spatial resolution 3D space (x, y, wavelength). Both $\alpha$ and $\beta$ are regularization parameters that tune the weight of each penalty function. In order to find the ${\mathbf{\hat{x}}}$ that minimizes Eq. \ref{eq:eq1}, we use a gradient descent algorithm. Given the gradient of the objective function:
\begin{equation}
\begin{split}
\nabla F(\mathbf{x}) &= T^\mathsf{T} S^\mathsf{T}(S T \mathbf{x} - \mathbf{y}_{cmos})
				+\alpha S^\mathsf{T} R_L^\mathsf{T} (R_L S \mathbf{x} - \mathbf{y}_{temporal}) +\\
				&+\beta T^\mathsf{T} R_L^\mathsf{T} (R_L T \mathbf{x} - \mathbf{y}_{spectral}),
\label{eq:eq3}
\end{split}
\end{equation}
we iteratively obtain ${\mathbf{\hat{x}}}$ by repeating ${\mathbf{\hat{x}}}_{n+1} = {\mathbf{\hat{x}}}_n - \tau \nabla F(\mathbf{\hat{x}}_n)$ until the solution converges \cite{repo_4d} (see Supplement for additional information and an outline of the code).

\begin{figure}[t]
\centering
\includegraphics[width=\linewidth]{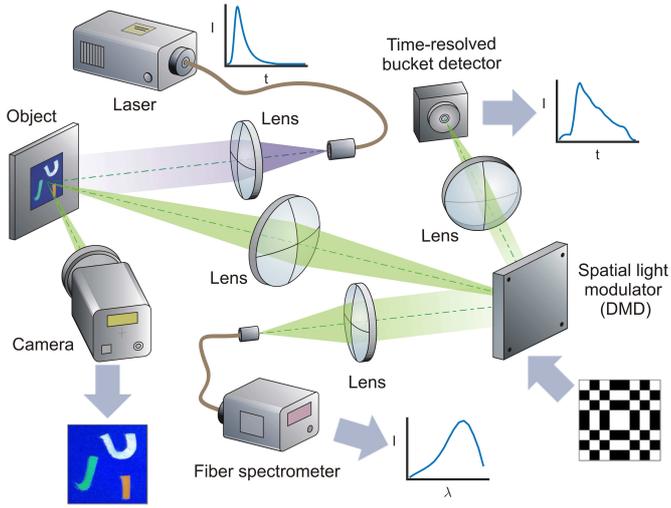}
\caption{ {\bf Optical implementation of the system.} The object is illuminated in reflectance geometry with a laser beam. The camera records a high-resolution 2D image of the object. An image of the object is also projected on the DMD. A sequence of Hadamard patterns is codified on the DMD at a high frame rate. For each pattern, the light emerging from the DMD is collected simultaneously by a time-resolved bucket detector and a spectrometer coupled with a detector array.}
\label{fig:fig2}
\end{figure}

A proposal for the experimental implementation of the system is shown in Fig. \ref{fig:fig2}. A 40~MHz pulsed supercontinuum laser source (Fianium, SC450) spectrally filtered through a band-pass filter (CW=480~nm,$\pm$5~nm), illuminates the sample under study, which consists of a plaque with three letters (U, J, and I). The U character contains the laser dye DCM, painted on a white paper, while the characters J and I are made of fluorescent plastic slides, respectively emitting in the green and orange region. The illumination area is $2.5 \times 2.5$ $\mathrm{cm}^2$. A CMOS camera is used to acquire an image of the sample over a single spectral band ($\mathbf{y}_{cmos}$). In parallel, a relay system images the sample onto the surface of a digital micromirror device (DMD, Discovery Kit 4100, Vialux). The DMD sequentially codifies the structured binary masks for SP image acquisition. In order to speed-up acquisition and to improve light efficiency, we use both reflection arms of the DMD in parallel. In one reflection direction, we place a time-resolved detector, which makes it possible to follow the temporal evolution of the fluorescence emission. In the other reflection direction, we combine a spectrometer with a detector array that allows to measure the different spectral components. After all the masks are generated by the DMD, the signal from each detector can be used to recover a low spatial resolution multispectral ($\mathbf{y}_{spectral})$ or time-resolved ($\mathbf{y}_{temporal}$) image by a simple multiplexing procedure that can easily be done on-the-fly. 

In our experiments, we acquired a $512 \times 512$ px image with the CMOS camera (Grasshopper3 GS3-U3-23S6M, Point Grey Research). The multispectral SP camera produced a $32 \times 32 \times 16$ datacube ($32 \times 32$ pixels with 16 spectral channels covering a range between 510 and 650 nm). It consisted of an imaging spectrometer (Acton, sp-2151i, Princeton Instruments) coupled to a 16-channel Photo-Multiplier Tube (PML16-C, Becker \& Hickl). The time-resolved SP camera is based on a Hybrid-PMT (HPM-100-50, Becker \& Hickl) connected to a Time-Correlated Single-Photon Counting board (TCSPC, SPC130EM, Becker \& Hickl) board, which is capable of providing photon time-of-flight histograms on a temporal window of about 25 ns. The overall data provided by the SP camera is a $32 \times 32 \times 256$ datacube ($32 \times 32$ pixels with 256 time bins of 48.8 ps each).

\begin{figure}[t]
\centering
\includegraphics[width=\linewidth]{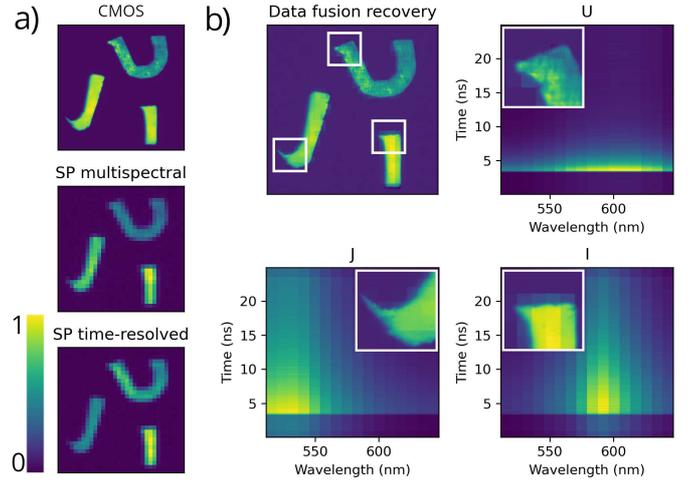}
\caption{ {\bf Time-resolved multispectral results.} a) Measured datasets. Top: CMOS image. Center: spatial projection of the multispectral SP datacube. Bottom: spatial projection of the time-resolved SP datacube. b) Spatial projection of the DF-recovered 4D hypercube and temporal-spectral traces for the different shapes present on the sample (labeled U, J, and I). Insets show the increased spatial resolution when compared to the SP datasets.}
\label{fig:fig3}
\end{figure}

Given the nature of SP imaging, both the multispectral and the time-resolved images share the same point of view of the scene. Nevertheless, the CMOS sees the scene under a different perspective. In order for the DF algorithm to work, we applied a pre-processing step that consisted on a spatial registration between the SP images and the CMOS image. This was performed using the Registration Estimator App (\textit{registrationEstimator}), available in Matlab. After the registration was done, a geometrical transformation was applied to the CMOS image in order to overlap its field of view with that of the SP images. The spatial projection of the results of each individual acquisition can be seen in Fig. \ref{fig:fig3}.a. After this procedure, the three datasets were fed to the DF algorithm, which produced a $512 \times 512 \times 16 \times 256 \approx 1$ giga-voxel hypercube. The complete reconstruction procedure consisted in 17 gradient descent steps, which took about 40 minutes. The computation was done using Matlab in a PC with an Intel Core i7-9700 CPU, with 64 Gb of RAM. A movie showing the individual temporal evolution of all the spectral channels can be found in Visualization 1.

Fig. \ref{fig:fig3}.b shows the DF recovery provided by fusing the three individual datasets. An increase in the spatial resolution of the images when compared to the SP measurements can be easily seen. While the improvement might not seem so high, acquiring $512 \times 512$ spatial resolution hypercubes only with the two SP systems would entail acquisition times 256 times longer (due to the sequential nature of SP imaging). We also show the temporal-spectral traces for different regions of the sample. In this visualization we can notice that the regions with the J and I characters present very similar fluorescence emission lifetimes, while the regions with the U and I characters have very similar spectral signatures. Exploiting both spectral and temporal information we can identify the 3 fluorescent species present in the sample. From the individual datasets alone, it would not be possible to do this classification.

\begin{figure}[ht!]
\centering
\includegraphics[width=\linewidth]{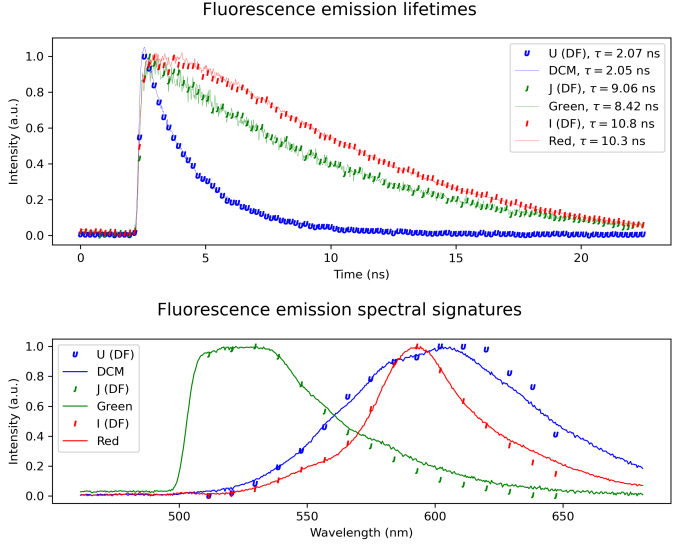}
\caption{ {\bf Temporal-spectral traces quality estimation.} Temporal (top) and spectral (bottom) traces for the three species present in the sample (U, J, and I characters). Lines correspond to the reference lifetimes and spectral signatures present in the sample, while the markers correspond to the values extracted from our 4D reconstruction. To ease visualization, we only show one of every two intensity values recovered by the DF algorithm in the emission lifetimes.}
\label{fig:fig4}
\end{figure}

In order to test the quality of our results, we compared the recovered spectra and fluorescence lifetimes with a reference of the species present in the sample. For the fluorescence lifetimes, we measured the decay time  of each fluorescent region with a fast detector (1024 temporal bins of 12.2 ps each). We show both the normalized data extracted from our DF reconstruction and the reference lifetimes in the top graph of Fig. \ref{fig:fig4}. From each one of the curves, it is possible to estimate the decay time by fitting the data to an exponential function. The values extracted from the DF reconstruction for the U, J, and I characters are $\tau_{U}^{DF}=2.07$ ns, $\tau_{J}^{DF}=9.06$ ns, and $\tau_{I}^{DF}=10.8$ ns, showing a very good agreement with the reference decays for the three fluorophores. Following the same spirit, we measured the fluorescence emission spectra for the three fluorophores in the scene using a high-resolution spectrometer (Hamamatsu TM-VIS/NIR C10083CA-2100), which also shown excellent agreement with the DF results.

In summary, we have introduced a novel DF-inspired multidimensional SP imaging system that can be used to identify different fluorescent species by their spectral and temporal signatures (i.e. their fluorescence spectra and/or emission lifetimes) and to study their photophysical properties. The system utilizes both array and SP detectors, combining their strengths while mitigating their drawbacks. In order to combine the individual datasets acquired by each camera, we have introduced a straightforward yet powerful DF recovery algorithm based on the minimization of a cost function that takes into account all the measurement processes. By doing so, we have demonstrated that it is possible to obtain high quality results in a fast manner while actually measuring a very small fraction of the information contained by the sample. In fact, if we consider the number of measured (M) vs. reconstructed (N) voxels for our experiments, we can think of the system as a compressive time-resolved multispectral camera, where the measurement ratio can be defined as $M.R.= M/N = \frac{512 \times 512 + 32 \times 32 \times 16 + 32 \times 32 \times 256 }{512 \times 512 \times 16 \times 256} \approx 0.0003$. In the future, we envision the use of more sophisticated cost functions introducing additional information of the system, such as sparsity constraints. This will further decrease the amount of measured information. While the results shown here consist of spatial-spectral-temporal information, the technique can be applied to any system consisting of multiple specialized cameras, and we expect that the DF paradigm will be useful for the bio-imaging community by also adding polarization and/or phase information.

\section*{Funding}
Ministerio de Ciencia e Innovación (PID2019--110927RB--I00 / AEI / 10.13039/501100011033); Generalitat Valenciana (PROMETEO/2020/029); Universitat Jaume I (UJIB2018--68); Regione Lombardia NEWMED, POR FESR 2014--2020.

\section*{Acknowledgments}
We acknowledge financial support from Laserlab--Europe (Grant Agreement n. 654148, Horizon 2020) through project CUSBO002482. A.J.M. Lenz acknowledges a grant from Generalitat Valenciana (ACIF/2019/019).

\section*{Disclosures} The authors declare no conflicts of interest.

\clearpage
\section*{Supplementary information}
\section{Data fusion retrieval algorithm}
As described in the main text, we model our system by using a forward model that contains three different terms, each one representing the measurements by each sensor. For each camera, the measurements ($\mathbf{y}_{cmos}$, $\mathbf{y}_{spectral}$, and $\mathbf{y}_{temporal}$) are obtained by projecting a 4D object into a 1D array of measurements. In order to implement this process, we define several routines in Matlab that integrate the 4D object into one or more dimensions (spectral, time) and/or either downsample the information in the spatial domain (as the single-pixel images are low-resolution versions of the true object). By using these forward operators, we define an objective function that can be minimized by using gradient descent.
In order to compute the gradient, we also implement several routines to calculate the adjoint of these operators \cite{claerbout2008basic}. All these, with a low-resolution example of our experiments, can be seen in \cite{repo_4d}.

The gradient descent procedure is also implemented in Matlab, with the only peculiarity of an intermediate step that searches for the best gradient step ($\tau$) at each iteration \cite{noauthor_backtracking_2021}. The pseudocode of the procedure can be seen in Alg. \ref{alg:a1}. Here, we tuned the regularization parameters empirically. However, for more complex forward models (for example including sparsity terms), this task could become extremely time consuming. Future experiments will explore the possibility to use automatic prediction of these paremeters \cite{Liao_Blind_deconv_2011, langer_automated_2017, liu_machine-learning-based_2021}, which would speed-up reconstruction process even with higher number of regularization terms.

\begin{algorithm2e}
\DontPrintSemicolon
\SetAlgoLined
\KwResult{Returns the estimated $\mathbf{\hat{x}}$ by minimizing the objective function $F(\mathbf{x})$ after a number $numIter$ of gradient descent steps. In each step an appropriate step size is calculated with a backtracking line search algorithmn based on the Armijo-Goldstein condition.}
Set values for the regularization parameters $\alpha$ and $\beta$, and for the initial step size $\tau_{init}$ (e.g., $\tau_{init}=1$)\;

Set values for the backtracking line search parameters $\varepsilon \in ( \, 0 \, , 1 )$ , and $\gamma \in ( \, 0 \, , 1 )$ (e.g., $\frac{1}{2}$ for both)\;

Initialize first guess $\mathbf{\hat{x}_{1}}$ randomly\;

\For{$i=1:numIter$}{
 Calculate gradient: $\mathbf{g_{i}}=\nabla F(\mathbf{\hat{x}_i})$\;

 Calculate new step size with a backtracking line search algorithm based on the Armijo-Goldstein condition:\;

Initialize step size: $\tau_{i}=\tau_{init}$\;

 \While{$F(\mathbf{\hat{x}_{i}})-F(\mathbf{\hat{x}_{i}}-\tau_{i} \, \mathbf{g_{i}}) < \varepsilon \, \tau_{i} \, \| \mathbf{g_{i}} \|^{2} $}
 {
  Set $\tau_{i} \leftarrow \gamma \tau_{i}$\;
 }
 Update object estimation in the direction provided by  the gradient:\;
 
 $\mathbf{\hat{x}_{i+1}} = \mathbf{\hat{x}_{i}} - \tau_{i} \mathbf{g_{i}}$\;
 
}

\caption{Gradient descent algorithm with backtracking line search}
\label{alg:a1}
\end{algorithm2e}

\clearpage
\bibliography{fusion_milano}

\end{document}